\documentclass[11pt,a4paper,twoside]{article}

\usepackage{titlesec}
\usepackage[shortlabels]{enumitem}
\titleformat{\section}
  {\normalfont\sffamily\fontsize{12}{15}\bfseries}{\thesection}{1em}{}
\titleformat{\subsection}
  {\normalfont\sffamily\fontsize{12}{15}\itshape}{\thesubsection}{1em}{}
\renewcommand{\abstract}{\textbf{Summary}.}
\makeatletter
\renewcommand{\maketitle}{\bgroup\setlength{\parindent}{0pt}
\begin{flushleft}
  \sffamily\textbf{\Large \@title}\\
  ~\\
  \@author\\
  \@date
\end{flushleft}\egroup
}
\makeatother

\usepackage[a4paper]{geometry}
\usepackage[utf8]{inputenc}
\usepackage{amsfonts}
\usepackage[hyphens]{url}
\usepackage{hyperref}
\hypersetup{
	colorlinks=true,
	linkcolor=black,
	citecolor=black,
	filecolor=black,
	urlcolor=black,
}
\usepackage{comment}
\usepackage{color}
\usepackage{microtype}
\usepackage{mathtools}
\usepackage{relsize}
\usepackage{fancyvrb}
\usepackage{amssymb,amsmath}
\usepackage{tikz}
\usepackage{subcaption}
\usepackage{natbib}

\newtheorem{definition}{Definition}
\newcommand{\+}[1]{\ensuremath{\mathbf{#1}}}
\newcommand{\po}[2]{#1(#2)}
\newcommand{\doo}{\textrm{do}}
\newcommand{\given}{{ \, | \, }}
\newcommand{\z}{trapdoor variable}

\usepackage{algorithm}
\usepackage{algorithmicx}
\usepackage[noend]{algpseudocode}

\usetikzlibrary{positioning, arrows.meta, shapes.geometric, patterns}
\tikzset{%
	thick/.style = {line width = 1pt},
	-{Latex},
	>={Latex},
	every edge/.append style={semithick},
	obs/.style = {name = #1, circle, draw, inner sep = 6pt, label = center:$#1$},
	clu/.style = {name = #1, regular polygon, regular polygon sides = 4, draw, inner sep = 6pt, label = center:$\+{#1}$}
}

\title{Estimation of causal effects with small data in the presence of trapdoor variables}
\author{Jouni Helske, Santtu Tikka and Juha Karvanen}
\date{\textit{Department of Mathematics and Statistics, University of Jyvaskyla, Finland.} \\
E-mail: {\scriptsize jouni.helske@jyu.fi}}

\begin{document}
\maketitle
	
\begin{abstract}
	We consider the problem of estimating causal effects of interventions from observational data when well-known back-door and front-door adjustments are not applicable. We show that when an identifiable causal effect is subject to an implicit functional constraint that is not deducible from conditional independence relations, the estimator of the causal effect can exhibit bias in small samples. This bias is related to variables that we call \emph{trapdoor variables}. We use simulated data to study different strategies to account for trapdoor variables and suggest how the related trapdoor bias might be minimized. The importance of trapdoor variables in causal effect estimation is illustrated with real data from the Life Course 1971--2002 study. Using this dataset, we estimate the causal effect of education on income in the Finnish context. Bayesian modelling allows us to take the parameter uncertainty into account and to present the estimated causal effects as posterior distributions. 
\end{abstract}

	
\section{Introduction}
\label{sec:introduction}

	Understanding causal relations forms the basis of decision making in society. The role of statistics is to provide tools that allow us to estimate the causal effects of planned interventions. Instead of modelling just associations, a causal model describes the functional relationships present in the system of interest. A core feature of causal models \citep{Pearl:book2009} is the capability to represent the effects of actions on the model via symbolic interventions. These interventions are assumed modular in the sense that they only modify the target of the intervention leaving other causal mechanisms in the model intact. The resulting probability distribution of this post-interventional causal model is defined as the causal effect of the intervention.
	
	Various methods have been suggested for estimating causal effects in different settings. Propensity score matching \citep{rosenbaum1983, imbens2000}, inverse probability weighting \citep{rosenbaum1983, rosenbaum1987}, and $g$-methods \citep{Robins1992} are some of the most well-known methods. These methods try to mimic a randomized experiment by creating pseudo-samples or by weighting the original sample in various ways, and it is typically assumed that there are no unobserved confounders present. In the classical structural equation modelling (SEM) approach \citep[see, e.g.,][]{Kline2011}, observed variables are assumed to be Gaussian, and unobserved confounders are treated as correlations between observed variables (there are also some extensions, such as LiNGAM for linear non-Gaussian cases \citep{shimizu2014}). 
	
	Another approach to causal inference is based on causal graphs, where the notion of identifiability plays a central role (see \autoref{sec:theory}). For certain graphs, the well-known back-door and front-door adjustment criteria \citep{Pearl1995} allow for identification of interventional distributions from observational data in the presence of unobserved confounders. The back-door criterion tells us whether a set of variables forms an admissible set, so that we only need to condition on these variables when estimating the causal effect, whereas the similar front-door criterion can be applied in the presence of a mediator between the interventional variable and the response variable. More generally, do-calculus \citep{Pearl1995} can be used to assess whether the interventional distribution of interest is identifiable given only the known causal graph, without any parametric assumptions about the distributions of the variables or the form of the effects they have on each other. Do-calculus provides an identifying functional, i.e., a nonparametric formula for the interventional distribution consisting of terms that represent observational distributions. Although identifiability does not in general guarantee estimability \citep{maclaren2019}, it is usually possible to obtain an estimator for the causal effect by replacing the terms present in the identifying functional by suitable parametric or nonparametric estimators and then combining the results accordingly. 
	
	We study the estimation of causal effects with small data in scenarios where standard adjustment criteria are not applicable. By small data, we refer to a case where parameter estimation exhibits non-negligible uncertainty due to the sample size. In addition, we consider the presence of functional equality constraints known as \emph{Verma-constraints} \citep{verma1990,TianPearl2002,robins1986}. Under certain conditions (see Section~\ref{sec:theory}), these constraints are related to special variables that we call \emph{trapdoor variables}. These variables can bias the causal effect estimator for finite samples and we refer to this form of bias as \emph{trapdoor bias}. We demonstrate the practical ramifications of trapdoor variables for the estimation of causal effects via simulations in a number of synthetic scenarios with small sample sizes and compare a variety of estimation strategies in both nonparametric and parametric settings. 

	As a motivating example, we consider Bayesian estimation of the causal effect of education on yearly income using real data from the Finnish Life Course 1971--2002 study \citep{fsd}. We construct a causal model for this study where we take into account the grade point average (GPA) from primary school, language skills, gender and the socioeconomic status of the parents. In the causal model, we find that GPA is in fact a trapdoor variable due to a functional equality constraint on the causal effect of interest. Bayesian modelling allows us to estimate the full post-interventional distribution of the income on different levels of education, which indicate a clear positive causal effect of education on income.
	We combine Bayesian estimation with a specialized Monte Carlo approach in order to take the effect of the trapdoor variable into account in a number of scenarios including the Life Course model. All analysis was done in the R environment \citep{R}, and the codes for the simulation experiments, the Life Course example, and the figures for the simulation results (created with the \texttt{ggplot2} package~\citep{ggplot2}) of this paper are available at \url{https://github.com/helske/trapdoor}.

	The paper is structured as follows. Section~\ref{sec:theory} introduces the notation, gives a definition for the trapdoor variables and present examples on causal models where such variables are present. Section~\ref{sec:estimation} focuses on various aspects related to the estimation of causal effects in the presence of trapdoor variables including a Bayesian approach and demonstrates how a trapdoor variables manifests under a linear-Gaussian model. Section~\ref{sec:simulations} considers the effect of trapdoor variables and the trapdoor bias of causal effect estimators via simulation in a model with binary variables, a linear-Gaussian model and a synthetic scenario based on the Life Course model. The analysis using the real Life Course data is presented in Section~\ref{sec:real-data-example}. Section~\ref{sec:discussion} provides some concluding remarks.
	
\section{Theory}
\label{sec:theory}
	
\subsection{Notation and basic definitions}

	Our analysis is based on the framework of \emph{structural causal models} (SCM) and directed graphs, and we assume the reader to be familiar with these concepts and their core probabilistic and graphical properties. For a more detailed discussion on SCMs and graph theoretic concepts, we refer the reader to works such as \citep{Pearl:book2009} and \citep{Koller09}. 
	
	We use capital letters to denote variables ($V$) and small letters to denote their values ($v$). Bold letters are used to denote sets of variables ($\+V$) and value assignments $(\+ v)$. The set of all possible value assignments to $\+ V$ is denoted by $val(\+ V)$. Set difference of sets $\+ A$ and $\+ B$ is denoted by $\+ A \setminus \+ B$. We use shorthand notation  $P(Y \given x)$ to denote the probability distribution $P_{\theta}(Y \given X = x)$ where we typically omit the dependence of (unknown) model parameters $\theta$.
	
	Each SCM $M$ over a set of variables $\+ V$ is associated a joint probability distribution $P(\+V)$ in a population of interest and a \emph{causal graph} $G$ over $\+ V$ where directed edges between two observed variables in $\+ V$ correspond to direct causal relationships which are assumed to not form any cycles. Bidirected edges between two observed variables in $\+ V$ are used to denote confounding by an unobserved common cause. In this framework, interventions are represented using the $\doo(\cdot)$-operator. An intervention $\doo(\+ X = \+x)$ forces the variables in $\+ X$ to take the values specified by $\+ x$ while leaving other mechanisms of the model intact. This intervention induces a submodel $M_{\+ x}$ with the interventional distribution $P(\+ V \given \doo(\+ X = \+ x))$. A causal effect $P(\+ Y \given \doo(\+ X = \+ x))$ is said to be \emph{identifiable} in $G$ if it is uniquely computable from $P(\+ V)$ in any SCM that induces $G$. A variable $Y$ in the post-intervention model $M_{\+ x}$ is denoted as $\po{Y}{\+ x}$. For an identifiable causal effect, an \emph{identifying functional} is a function $f$ such that $f(P(\+ V)) = P(\+ Y \given \doo(\+ X = \+ x))$. We assume that $P(\+ V = \+ v) > 0$ for values $\+ v \in val(\+ V)$ making all conditional distributions and interventions well-defined. 
	
	Note that identifiability only indicates the existence of an estimator and does not take into account the potential problems stemming from finite data. Therefore, even though determining the identifiability of $P(\+ Y \given \doo(\+ X = \+ x))$ is an important first step, it does not guarantee that we can estimate the causal effect in practice without additional stability assumptions \citep{maclaren2019}. Note also that through this paper we assume that our causal graph is correct, while in practice we are rarely certain of this.
	
\begin{figure}[!ht]
		\centering
		\begin{tikzpicture}[scale = 0.6]
		\node [obs = {Y}] at (6,0) {$\vphantom{X}$};
		\node [obs = {X}] at (2,0) {$\vphantom{X}$};
		\node [obs = {Z}] at (-2,0) {$\vphantom{X}$};
		\node [obs = {W}] at (-6,0) {$\vphantom{X}$};
		\path [->] (X) edge (Y);
		\path [->] (Z) edge (X);
		\path [->] (W) edge (Z);
		\path [<->,dashed] (Z) edge [bend left = 30] (Y);
		\end{tikzpicture}
		\caption{A causal graph where the identifying functional of $P(Y \given \doo(X = x))$ obtained by an application of the ID algorithm does not depend on the value of of $W$ and there is an admissible set $Z$ enabling back-door adjustment.}
		\label{fig:verma-admissible}
	\end{figure}
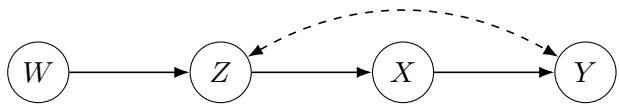

	Interventional distributions exhibit conditional independence constraints, which can be characterized by $d$-separation \citep{Pearl88} in the associated causal graph of the model. However, identifiable causal effects can be subject to Verma-constraints. As an example of such a constraint,  we show that in the causal graph of \autoref{fig:verma-admissible},
	a causal effect does not depend on the value of a variable $W$ despite it appearing in the identifying functional of the interventional distribution. The causal effect of $X$ on $Y$ is identifiable in this graph which can be verified using do-calculus or by applying an identifiability algorithm such as the one by \citet{huangvaltorta:complete} or the ID algorithm by \citet{Shpitser2006}. Application of the ID algorithm implemented in the \texttt{R} package \texttt{causaleffect} \citep{TikkaKarvanen2017} provides the formula
	\[ P(Y \given \doo(X = x)) = \sum_{z} P(Y \given x, z, w) P(z \given w), \]
	where the left-hand side depends on the value of $X$ and $Y$, but on the right-hand the variable $W$ is also present and it is not subject to summation, unlike the variable $Z$. However, the right-hand side cannot depend on the value of $W$, as there is an admissible set $Z$ which gives us an alternative formula
	\begin{equation*} 
	P(Y \given \doo(X = x)) = \sum_{z}P(Y \given x, z)P(z),
	\end{equation*}
	i.e., a simple back-door formula where variable $W$ is not present.
	
	Perhaps surprisingly, Verma-constraints can be expressed as conditional independences in the interventional distribution \citep{ShpitserPearlDormant} and can be used to give an alternative definition for nested Markov models in acyclic directed mixed graphs \citep{richardson2012}. 
	Verma-constraints have been used for testing edges \citep{Shpitser2009} and for marginalization via variable elimination \citep{Shpitser2011}.
	
\subsection{Trapdoor variables}\label{sec:trapdoor}

	Here we give a broad definition that captures the notion that a set of variables $\+ Z$ may appear in an identifying functional of a causal effect, but the value of the causal effect is not dependent on the value of $\+ Z$. 
	\begin{definition}[Functional independence]
		Let $\+ X, \+Y, \+ Z \subset \+ V$ be disjoint sets and let ${P(\+Y \given \doo(\+ X  = \+ x))}$ be an identifiable causal effect with an identifying functional $g(\+ v) = f(P(\+ v))$. If the domain of $g$ is $val(\+ X) \times val(\+ Y) \times val(\+ Z)$, and $g(\+ x, \+ y, \+ z_1) = g(\+ x, \+ y, \+ z_2)$ for all $\+ x \in val(\+ X)$, $\+y \in val(\+ Y)$ and $\+ z_1,\+ z_2 \in val(\+ Z)$, then $g$ is \emph{functionally independent} from $\+ Z$. 
	\end{definition}
	The constraint defined above is specific to the given identifying functional. In some instances we may be able to find identifying functionals that do not exhibit functional equality constraints for any subset $\+ Z$ of $\+ V$. Our interest lies in the opposite direction, where every identifying functional exhibits a specific type of functional independence. Before characterizing this property of interest, we must first define an operation known as the latent projection of a causal graph \citep{pearl1991}.
	\begin{definition}[Latent projection] Let $G$ be a causal graph over a set of vertices $\+ V \cup \+ L$. The \emph{latent projection} $L(G, \+ V)$ is a causal graph over $\+ V$ where for every pair of distinct vertices $Z,W \in \+ V$ it holds that
		\begin{enumerate}[(a),leftmargin=*]
			\item $L(G, \+ V)$ contains an edge $Z \longrightarrow W$ if there exists a directed path $Z \longrightarrow \cdots \longrightarrow W$ in $G$ on which every vertex except $Z$ and $W$ is in $\+ L$.
			\item $L(G, \+ V)$ contains an edge $Z \longleftrightarrow W$ if there exists a path from $Z$ to $W$ in $G$ that does not contain the pattern $Z \longrightarrow M \longleftarrow W$ (a collider), on which every vertex except $Z$ and $W$ is in $\+ L$, the first edge of the path has as arrowhead into $W$, and the last edge has an arrowhead into $Z$.
		\end{enumerate}
	\end{definition}
	Latent projections can be used to derive identifying functionals for causal effects such that they do not contain a specific variable, i.e., the variable is considered latent, and the causal effect of interest is identified in the corresponding latent projection \citep{tikka18}. For this reason the presence of a functional constraint in some identifying functional does not rule out the possibility of obtaining another identifying functional that is not subject to the same constraint (Recall the example on the identifying functional of $P(Y \given \doo(X = x))$ in the graph of \autoref{fig:verma-admissible}).

	We rule out this possibility of finding alternative identifying functionals by restricting our attention to settings where a \emph{trapdoor variable} is present.
	\begin{definition}[Trapdoor variable] If ${P(\+ Y \given \doo(\+ X = \+ x))}$ is identifiable in $G$ from $P(\+ V)$, its identifying functional $f(P(\+ V))$ is functionally independent of $\+ Z$ where $\+ Z \subset \+ V \setminus (\+ X \cup \+ Y)$, and $P(\+ Y \given \doo(\+ X = \+ x))$ is not identifiable in $L(G, \+ V \setminus \+ Z)$ from $P(\+ V \setminus \+ Z)$, then $\+ Z$ is a set of \emph{\z s} with respect to $f(P(\+ V))$ in $G$.
	\end{definition}
	Finding \z s is straightforward as outlined by their definition. Given a causal effect of interest, we first determine its identifiability and whether the identifying functional is subject to functional equality constraints. If this is the case, we proceed to verify whether the causal effect might be identifiable when the set $\+ Z$ is considered latent using a suitable latent projection. The algorithm by \citet{TianPearl2002} can be used to systematically enumerate functional equality constraints implied by a causal model. \citet{evans2018} showed that this constraint-finding algorithm is complete for categorical variables, but it is not known whether the algorithm is complete in general, i.e., whether all such constraints can be found by the algorithm.
	
	If there exists a trapdoor variable for an identifying functional of a causal effect of interest, then the estimate of the causal effect may depend on the value of the trapdoor variable. This dependency can introduce bias in the estimate.
	\begin{definition}[Trapdoor bias]
		Let $\hat g(\+ v)$ be an estimator of an identifying functional $g(\+ v)$ of a causal effect $P(\+ Y \given \doo(\+ X  = \+ x))$ and let $\+ Z$ be a set of trapdoor variables with respect to $g(\+ v)$. Let $B(\hat g(\+ v))$ denote the bias of this estimator. If there exists $\+ z_1, \+ z_2 \in val(\+ Z)$ so that $B(\hat g(\+ x,\+ y,\+z_1)) \neq B(\hat g(\+x, \+ y, \+ z_2))$, then $\hat g(\+ v)$ exhibits \emph{trapdoor bias} with respect to $\+ Z$.
	\end{definition}
	For a consistent estimator, the effect of the trapdoor bias becomes negligible as the sample size grows, but for small samples the choice of how to take the trapdoor variables into account may be significant. 
	
\subsection{Example on trapdoor variables}
\label{sec:verma3}
	\begin{figure}[!ht]
		\centering
		\begin{subfigure}{0.32\textwidth}
			\centering
			\begin{tikzpicture}[scale=0.9]
			\node [obs = {W}] at (-2,1) {$\vphantom{X}$};
			\node [obs = {Y}] at (1,-2) {$\vphantom{X}$};
			\node [obs = {X}] at (-2,-2) {$\vphantom{X}$};
			\path [->] (W) edge (X);
			\path [->] (X) edge (Y);
			\path [<->,dashed] (Y) edge [bend right = 45] (W);
			\end{tikzpicture}
			\caption{}
			\label{fig:dagA}
		\end{subfigure}
		%
		\begin{subfigure}{0.33\textwidth}
			\centering
			\begin{tikzpicture}[scale=0.9]
			\node [obs = {W}] at (-2,1) {$\vphantom{X}$};
			\node [obs = {Y}] at (1,-2) {$\vphantom{X}$};
			\node [obs = {X}] at (-2,-2) {$\vphantom{X}$};
			\path [->] (W) edge (X);
			\path [->] (X) edge (Y);
			\path [<->,dashed] (Y) edge [bend right = 45] (W);
			\path [<->,dashed] (W) edge [bend right = 45] (X);
			\end{tikzpicture}
			\caption{}
			\label{fig:dagB}
		\end{subfigure}
		%
		\begin{subfigure}{0.32\textwidth}
			\centering
			\begin{tikzpicture}[scale=0.9]
			\node [obs = {W}] at (-2,1) {$\vphantom{X}$};
			\node [obs = {Y}] at (1,-2) {$\vphantom{X}$};
			\node [obs = {X}] at (-2,-2) {$\vphantom{X}$};
			\node [obs = {Z}] at (-2,-0.5) {$\vphantom{X}$};
			\path [->] (X) edge (Y);
			\path [->] (Z) edge (X);
			\path [->] (W) edge (Z);
			\path [<->,dashed] (Y) edge [bend right = 45] (W);
			\path [<->,dashed] (W) edge [bend right = 45] (X);
			\end{tikzpicture}
			\caption{}
			\label{fig:dagC}
		\end{subfigure}
		\caption{Three causal graphs of increasing complexity, where the interest is in the causal effect of $X$ on $Y$.}
		\label{fig:dags}
	\end{figure}
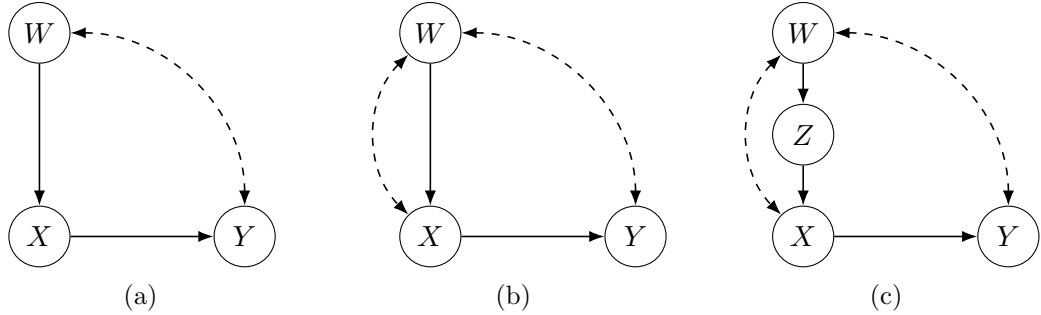
	\noindent
	Consider the three causal graphs in \autoref{fig:dags}. We are interested in estimating the causal effect of $X$ on $Y$. For example, in our application in \autoref{sec:real-data-example}, $X$ will be the education level and $Y$ is the yearly income. Furthermore, $Z$ corresponds to the GPA from primary school, and $W$ to the socioeconomic status of the parents. In all three graphs we have an arrow from $X$ to $Y$, meaning that we assume that there is a direct causal effect of education on income. In addition, we assume that there may be some unobserved confounders between socioeconomic status of the parents and income. In the graphs of \autoref{fig:dagB} and \autoref{fig:dagC}, we also assume that there is confounding between socioeconomic status of the parents and education level of the participant. In \autoref{fig:dagC}, we assume that the effect of socioeconomic status on the education level is mediated by  the GPA. We will further extended the third graph with additional variables, such as gender, in Section~\ref{sec:real-data-example}.
	
	Now consider the estimation of the interventional distribution $P(Y \given \doo(X = x))$, i.e., the distribution of $Y$ when we intervene on $X$ by setting it to $x$, which differs from a simple conditional distribution of $P(Y \given X = x)$ in these graphs. In order to estimate $P(Y \given \doo(X = x))$, we need to find a formula for it in terms of the observed variables only, i.e., $Y$, $X$, $W$, and $Z$ in our example. In the graph of \autoref{fig:dagA} we obtain the so called back-door adjustment formula
	\begin{equation*}
	P(Y \given \doo(X = x)) = \sum_{w}P(Y\given x, w)P(w).
	\end{equation*}
	
	By conditioning on $W$ we block all back-door paths from $X$ to $Y$ i.e., those paths between $X$ and $Y$ which have arrows into $X$. Thus by estimating the parameters of the terms $P(Y \given w, x)$ and $P(W)$ we can estimate the interventional distribution of interest, $P(Y \given \doo(X = x))$. For example, assuming that all variables are Gaussian and their relationships are linear, we have $\po{Y}{x} \sim N(a_y + b_{yw} a_w + b_{yx} x, s^2_y + b_{yw}^2 s^2_w)$, with $b_{ij}$ denoting the estimated regression coefficient of variable $j$ on variable $i$, $a_i$ denoting the intercept term and $s^2_i$ corresponding to the estimated residual variance.
	
	Now consider the graph of \autoref{fig:dagB}. In this case, conditioning on $W$ blocks the path $Y \longleftrightarrow W \longrightarrow X$ but opens the path $Y \longleftrightarrow  W \longleftrightarrow X$ meaning that we cannot apply the back-door adjustment here. In fact, adding the unobserved confounder between $W$ and $X$ renders the causal effect nonidentifiable.
	
	In the graph of \autoref{fig:dagC}, we assume that we have obtained data on variable $Z$ which lies on the directed path from $W$ to $X$. Given this additional information, the causal effect $P(Y \given \doo(X = x))$ is again identifiable, and we have 
	\begin{equation}
	\label{eq:dag3}
	P(Y \given \doo(X = x)) = \frac{\sum_{w}P(Y\given x, z, w)P(x \given z, w)P(w)}{\sum_{w}P(x \given z, w)P(w)}.
	\end{equation}
	However, there is no term for the distribution of $Z$ in equation~\eqref{eq:dag3}. Thus after estimating the distributions $P(Y \given  x, z, w)$, $P(X \given z, w)$, and $P(W)$, $Z$ is essentially reduced to a fixed but unknown parameter in the context of $P(Y \given \doo(X = x))$. This graph is also considered by \citet{TianPearl2002}, \citet{why}, and \citet{jung2020}. \citet{TianPearl2002} show that this graph contains a functional equality constraint of an interventional distribution which cannot be expressed as a conditional independence constraint using the observed variables. The constraint states that the formula for $P(Y \given \doo(X = x))$ given in equation~\eqref{eq:dag3} is functionally independent of $Z$, meaning that its value is independent on the choice of $z$, just as we would intuitively expect. However, when estimating equation~\eqref{eq:dag3} from the data we clearly must choose some value for $Z$, even though the constraint states that the actual value should not matter. In fact, $Z$ is a \z{} with respect to the identifying functional of equation~\eqref{eq:dag3} in this graph. This follows from the functional equality constraint and the fact that the causal effect is not identifiable in the causal graph of \autoref{fig:dagB} which is the latent projection of \autoref{fig:dagC} when $Z$ is considered latent. 
	
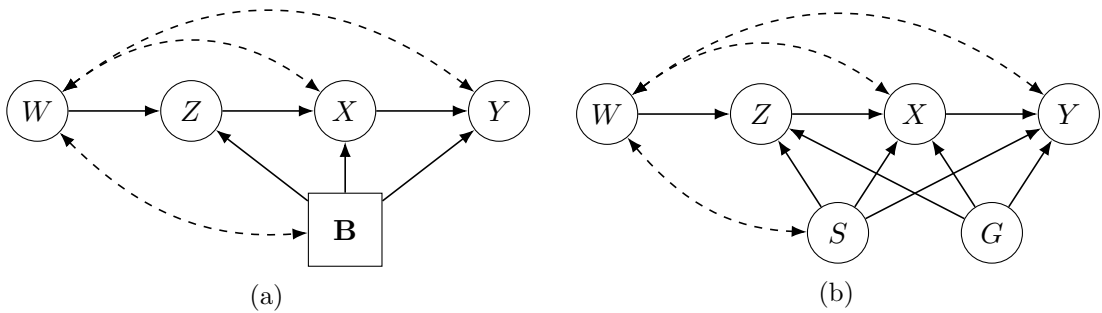
\begin{figure}[!ht]
		\centering
		\begin{subfigure}{0.49\textwidth}
			\centering
			\begin{tikzpicture}[xscale=0.75,scale = 0.9]
			\node [clu = {B}] at (0,-1.75) {$\vphantom{X}$};
			\node [obs = {Y}] at (3,0) {$\vphantom{X}$};
			\node [obs = {X}] at (0,0) {$\vphantom{X}$};
			\node [obs = {Z}] at (-3,0) {$\vphantom{X}$};
			\node [obs = {W}] at (-6,0) {$\vphantom{X}$};
			\path [->] (W) edge (Z);
			\path [->] (X) edge (Y);
			\path [->] (B) edge (Y);
			\path [->] (B) edge (X);
			\path [->] (B) edge (Z);
			\path [->] (Z) edge (X);
			\path [<->,dashed] (W) edge [bend right = 20] (B);
			\path [<->,dashed] (W) edge [bend left = 30] (X);
			\path [<->,dashed] (W) edge [bend left = 30] (Y);
			\end{tikzpicture}
			\caption{}
			\label{fig:general}
		\end{subfigure}
		\hfill
		\begin{subfigure}{0.49\textwidth}
			\centering
			\begin{tikzpicture}[xscale=0.75,scale = 0.9]
			\node [obs = {G}] at (1.5,-1.75) {$\vphantom{X}$};
			\node [obs = {S}] at (-1.5,-1.75) {$\vphantom{X}$};
			\node [obs = {Y}] at (3,0) {$\vphantom{X}$};
			\node [obs = {X}] at (0,0) {$\vphantom{X}$};
			\node [obs = {Z}] at (-3,0) {$\vphantom{X}$};
			\node [obs = {W}] at (-6,0) {$\vphantom{X}$};
			\path [->] (S) edge (Y);
			\path [->] (W) edge (Z);
			\path [->] (S) edge (X);
			\path [->] (S) edge (Z);
			\path [->] (X) edge (Y);
			\path [->] (G) edge (Y);
			\path [->] (G) edge (X);
			\path [->] (G) edge (Z);
			\path [->] (Z) edge (X);
			\path [<->,dashed] (W) edge [bend right = 20] (S);
			\path [<->,dashed] (W) edge [bend left = 30] (X);
			\path [<->,dashed] (W) edge [bend left = 30] (Y);
			\end{tikzpicture}
			\caption{}
			\label{fig:fsd}
		\end{subfigure}
		\caption{A generalization of the setting presented in \autoref{fig:dagC} with an additional set of variables $\+ B$ is shown in (a). The square node for $\+ B$ denotes an arbitrary causal graph over $\+ B$. Edges to and from $\+ B$ mean that such an edge may exist between any member of $\+ B$ and the other endpoint of the edge. An instance of (a) is shown in (b) with $\+ B = \{S,G\}$.}
	\end{figure}

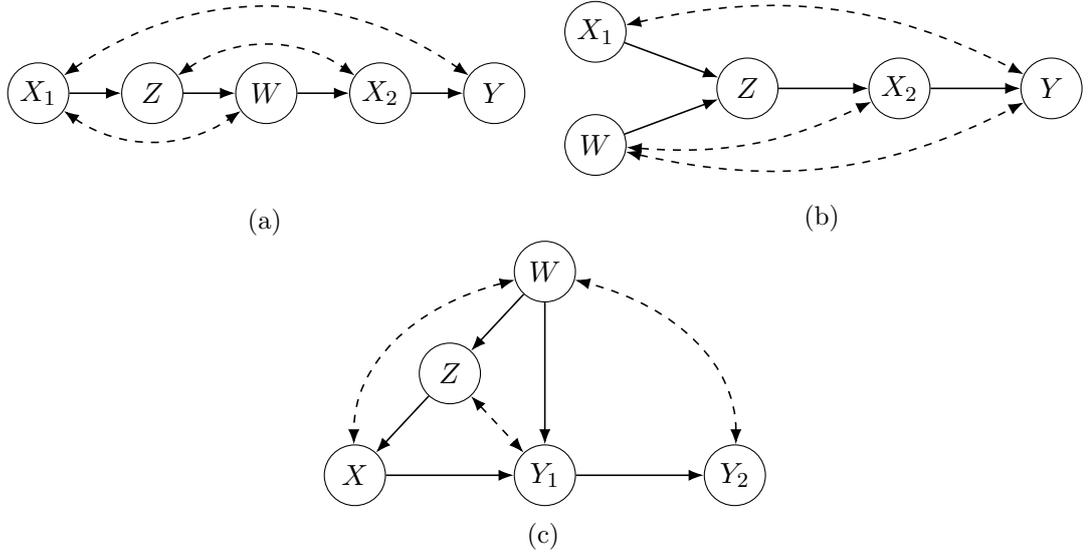
\begin{figure}[!ht]
		\centering
		\begin{subfigure}{0.49\textwidth}
			\centering
			\begin{tikzpicture}[scale=0.5]
			\node [obs = {Y}] at (6,0) {$\vphantom{X}$};
			\node [obs = {X_2}] at (3,0) {$\vphantom{X}$};
			\node [obs = {W}] at (0,0) {$\vphantom{X}$};
			\node [obs = {Z}] at (-3,0) {$\vphantom{X}$};
			\node [obs = {X_1}] at (-6,0) {$\vphantom{X}$};
			\node at (0,-2.1) {$\vphantom{X}$};
			\path [->] (X_2) edge (Y);
			\path [->] (W) edge (X_2);
			\path [->] (Z) edge (W);
			\path [->] (X_1) edge (Z);
			\path [<->,dashed] (X_1) edge [bend right = 35] (W);
			\path [<->,dashed] (X_1) edge [bend left = 35] (Y);
			\path [<->,dashed] (Z) edge [bend left = 35] (X_2);
			\end{tikzpicture}
			\caption{}
			\label{fig:extraA}
		\end{subfigure}
		\begin{subfigure}{0.49\textwidth}
			\centering
			\begin{tikzpicture}[scale=0.5]
			\node [obs = {Y}] at (6,0) {$\vphantom{X}$};
			\node [obs = {X_2}] at (2,0) {$\vphantom{X}$};
			\node [obs = {W}] at (-6,-1.5) {$\vphantom{X}$};
			\node [obs = {Z}] at (-2,0) {$\vphantom{X}$};
			\node [obs = {X_1}] at (-6,1.5) {$\vphantom{X}$};
			\path [->] (X_2) edge (Y);
			\path [->] (Z) edge (X_2);
			\path [->] (W) edge (Z);
			\path [->] (X_1) edge (Z);
			\path [<->,dashed] (W) edge [bend right = 20] (Y);
			\path [<->,dashed] (X_1) edge [bend left = 20] (Y);
			\path [<->,dashed] (W) edge [bend right = 15] (X_2);
			\end{tikzpicture}
			\caption{}
			\label{fig:extraB}
		\end{subfigure}
		\begin{subfigure}{0.49\textwidth}
			\centering
			\begin{tikzpicture}[yscale=0.9,xscale=1.25]
			\node [obs = {W}] at (0,1) {$\vphantom{X}$};
			\node [obs = {Y_2}] at (2,-2) {$\vphantom{X}$};
			\node [obs = {Y_1}] at (0,-2) {$\vphantom{X}$};
			\node [obs = {X}] at (-2,-2) {$\vphantom{X}$};
			\node [obs = {Z}] at (-1,-0.5) {$\vphantom{X}$};
			\path [->] (X) edge (Y_1);
			\path [->] (Y_1) edge (Y_2);
			\path [->] (Z) edge (X);
			\path [->] (W) edge (Z);
			\path [->] (W) edge (Y_1);
			\path [<->,dashed] (Y_2) edge [bend right = 35] (W);
			\path [<->,dashed] (W) edge [bend right = 35] (X);
			\path [<->,dashed] (Z) edge (Y_1);
			\end{tikzpicture}
			\caption{}
			\label{fig:extraC}
		\end{subfigure}
		\caption{Examples on graphs where trapdoor variables are present. In (a) and (b), $Z$ is a trapdoor variable for $P(Y \given \doo(X_1 = x_1,X_2 = x_2))$. In (c), $Z$ is a trapdoor variable for $P(Y_1,Y_2 \given \doo(X = x))$.}
		\label{fig:extras}
	\end{figure}

	Trapdoor variables are in no way limited to the scenario of \autoref{fig:dagC}. \autoref{fig:general} presents a generalization of \autoref{fig:dagC} with an additional set of variables $\+ B = \{B_1,\ldots,B_k\}$, that are possibly connected to each other with directed arrows or through unobserved confounders.  In addition, any member of $\+B$ is possibly connected to $W$ via a bidirected edge or to $Z$, $X$ or $Y$ via a directed edge. \autoref{fig:extras} depicts three additional examples of graphs where trapdoor variables are present. In the following sections we investigate the challenges that \z s impose on the estimation of causal effects. 

	\noindent

\section{Estimation}
\label{sec:estimation}

	\subsection{Plug-in estimation}\label{sec:plugin}
	A straightforward strategy for causal effect estimation given the assumed causal graph is to construct a parametric model for the conditional distributions that appear in the formula of an identifying functional of a causal effect. As an example, consider the graph of \autoref{fig:dagC}, and assume that we wish to estimate $E(Y \given \doo(X = x))$. We need to estimate the unknown model parameters $\theta$ corresponding to the causal effect of equation~\eqref{eq:dag3} and we let $P_{\hat\theta}(\cdot)$ denote the density where the unknown parameters $\theta$ are assigned some fixed values such as their maximum likelihood (ML) estimates. The plug-in estimator for the expected value of the interventional distribution takes the following form 
	\begin{equation} 
	\label{eq:plugin1}
	\widehat E_{\hat \theta, z}(Y \given \doo(X = x)) = \sum_{y} y \frac{\sum_{w}P_{\hat \theta}(y \given w, z, x)P_{\hat \theta}(x \given  w, z) P_{\hat \theta}(w)}{\sum_{w}P_{\hat \theta}(x \given  w, z) P_{\hat \theta}(w)}.
	\end{equation}
	Note that this estimate may depend on the value of $z$ when using a small sample estimate for $\theta$. This dependency and the corresponding trapdoor bias vanishes when the models are correctly specified and the true parameter values are used (by the definition of functional independence). However, even if we assume an unbiased estimator $\hat \theta$, the estimator~\eqref{eq:plugin1} can still exhibit bias since it is a nonlinear function of $\hat \theta$ that depends on the choice of $z$. For large enough datasets, the choice of $z$ in equation~\eqref{eq:plugin1} should not have a large impact, but we will show that for small samples, the value of the \z{} $z$ can have a crucial role in estimating the correct causal effect of $X$ on $Y$. 
	
	An alternative strategy for defining the plug-in estimator is to eliminate the functional constraint by averaging the identifying functional over a (conditional) distribution of the trapdoor variable. If $Z$ is a trapdoor variable and $\+ T \subseteq (\+ V \setminus Z)$, then
	\[
	P(Y \given \doo(X = x)) = f(P(\+ v)) \cdot 1 = f(P(\+ v)) \sum_{z} P(z \given \+ t) = \sum_{z} P(z \given \+ t) f(P(\+ v)),
	\]
	since $f(P(\+ v))$ is functionally independent of $Z$. This leads to the following alternative estimator for the expected value of the interventional distribution in the causal graph of \autoref{fig:dagC}
	\begin{equation} \label{eq:plugin2}
	\widehat E_{\hat \theta}(Y \given \doo(X = x)) = \sum_{y}y\sum_{z} P_{\hat \theta}(z \given \+ t) \frac{\sum_{w}P_{\hat \theta}(y \given w, z, x)P_{\hat \theta}(x \given  w, z) P_{\hat \theta}(w)}{\sum_{w}P_{\hat \theta}(x \given  w, z) P_{\hat \theta}(w)}.
	\end{equation}

	Note that while this strategy eliminates the problem of having to choose a specific value for $z$, it also introduces a new problem of selecting the set $\+ T$ and estimating the model parameters for the distribution $P_{\theta}(Z \given \+ t)$. As before, this estimator is a nonlinear function of the parameters that can result in bias. Note that in a frequentist framework, there is no way to divide the total bias of the final estimate into trapdoor bias, plug-in bias, or other possible sources of bias.
	
	\subsection{Bayesian estimation of causal effects}
	
	We advocate the use of Bayesian methods for jointly estimating the parameters of the distributions of identifying functionals of interest. By drawing samples from the joint posterior of the model parameters (using any generic Bayesian inference engine, typically some type of Markov chain Monte Carlo (MCMC) algorithm), we propagate the parameter estimation uncertainty to the final causal effect estimates and avoid the plug-in bias due to the nonlinear formula of the identifying functional with respect to model parameters. This allows us to focus on the effects of the trapdoor variable and the trapdoor bias of the estimators. We can also obtain samples from the full posterior of $P(Y \given \doo(X = x))$ which can be used for straightforward evaluation of any properties of interest (such as mean and variance) of this posterior. 
	
	In parametric estimation of causal effects, we typically do not have an analytical formula for the types of plug-in estimators that are shown in equations~\eqref{eq:plugin1} and \eqref{eq:plugin2}. In these cases, we can use a Monte Carlo approach to draw samples from $P_{\hat \theta}(Y \given \doo(X = x))$, and estimate the desired quantity using these samples. The specific Monte Carlo algorithm depends on the causal graph, the identifying functional and the corresponding conditional distributions. In simple cases, we simulate variables from their (conditional) distributions with fixed $x$, whereas for example in the case of equation~\eqref{eq:dag3} where $P(X \given \cdot)$ is present, we need additional weighting of the samples. Consider our extended example graph in \autoref{fig:general}, which leads to the following formula for $P(Y \given \doo(X = x))$:
		\begin{equation}
	\label{eq:fsd_general}
	 \sum_{\+b} P(\+b)\frac{\sum_{w} P(Y \given x, z, w, \+b)P(x \given z,w,\+b)P(\+b \given w)P(w)}{\sum_{w} P(x \given z,w,\+b)P(\+b\given w)P(w)}.
	\end{equation}
	Algorithm \ref{algo:mc} describes the Monte Carlo algorithm for equation \eqref{eq:fsd_general} based on the second approach discussed in \autoref{sec:plugin}, given the parameter vector $\hat \theta$ (the same algorithm is also suitable for \eqref{eq:dag3} by omission of the references to the variables in $\+ B$). Algorithm \ref{algo:mc} draws samples from the marginal and conditional distributions defined in \eqref{eq:fsd_general} and gives us $N \times M$ weighted replications from $P_{\hat \theta}(Y \given \doo(X = x))$ which allows us to compute, e.g.,
\begin{equation}\label{eq:mc-mean}
\widehat E_{\hat \theta}(Y \given \doo(X = x)) = \sum_{i=1}^{N}\sum_{j=1}^{M} \bar \gamma^{ij} y^{ij},
\end{equation}
where the weights $\bar \gamma^{ij}$ are defined in the last step of Algorithm~\ref{algo:mc}. 

\begin{algorithm}[!t]
	\begin{algorithmic}[1]
		\State For $i = 1,\ldots,N$:
		\State \quad Sample $\+b^i \sim P_{\hat \theta}(\+B)$
		\State \quad Set or sample the value of the \z, for example as $z^i = E_{\hat \theta}(Z)$ or $z^i \sim P_{\hat \theta}(Z \given  x, \+b^i)$
		\State \quad For $j = 1, \ldots, M$:
		\State \quad \quad Sample $w^{ij} \sim P_{\hat \theta}(W)$
		\State \quad \quad Sample $y^{ij} \sim P_{\hat \theta}(Y \given  x, z^i, w^{ij}, \+b^i)$
		\State \quad \quad Compute $\gamma^{ij} =  P_{\hat \theta}(x \given z^i, w^{ij}, \+b^i) P_{\hat \theta}(\+b^i \given  w^{ij})$
		\State Compute the normalized weights:
		\[
		\bar \gamma^{ij} = \frac{\gamma^{ij}}{\frac{1}{M}\sum_{i=1}^N \gamma^{ij}}, \quad i = 1, \ldots, N,\, j = 1, \ldots, M
		\]
	\end{algorithmic}
	\caption{Monte Carlo algorithm for sampling from $P_{\hat \theta}(Y \given \doo(X = x))$ defined by equation~\eqref{eq:fsd_general} with $N \times M$ Monte Carlo samples.}
	\label{algo:mc}
\end{algorithm}

Here the trapdoor variable has a more crucial role than in the case of analytical formulas. With poor choices of $z$, our weights $\bar \gamma^{ij}$ can become degenerate, i.e., most of the weights are near zero. In order to avoid this, it is natural to condition the trapdoor variable on $x$ and perhaps other variables such as members of $\+ b^i$ in Algorithm~\ref{algo:mc} which should make the weights well behaved.

The suitable number of Monte Carlo samples $N \times M$ can determined by computing the Monte Carlo standard error (MCSE) of our causal effect estimate. The MCSE measures the additional uncertainty in our estimate due to the finite Monte Carlo sample size. For example, the MCSE for estimator~\eqref{eq:mc-mean} can be computed as

\begin{equation*}
MCSE\left(\widehat E_{\hat \theta}(Y \given \doo(X = x))\right) = \sqrt{\sum_{i=1}^N\sum_{j=1}^M\left[\bar \gamma^{ij} \left\{y^{ij} - \widehat E_{\hat \theta}(Y \given \doo(X = x))\right\}\right]^2}.
\end{equation*} 
When combining Algorithm~\ref{algo:mc} with Bayesian parameter estimation, the algorithm is used at each MCMC iteration given the current values of the model parameters $\theta$. 
	
With the MCMC approach, we can compute functions of $\po{Y}{x}$ at each iteration, giving us samples from its posterior distribution, or we can store all $N \times M$ weighted Monte Carlo samples leading to posterior distribution of $\po{Y}{x}$, i.e., the variable $Y$ in the post-interventional distribution, which can be further used to evaluate functions of interest. In the latter case, it can be practical to resample (using the corresponding weights) and store only one replication $y^{ij}$ at each iteration if memory constraints limit the storing of all samples.

	\subsection{Analytical causal effect for a linear-Gaussian model}
	\label{sec:linear_gaussian_analytical}
	We will now consider the estimation of causal effects in a common linear-Gaussian case for the causal graph of \autoref{fig:dagC}. We assume that the underlying model is defined as
	\begin{equation} 
	\label{eq:dgm}
	\begin{aligned}
	U &\sim N(\mu_{u}, \sigma_{u}^2),\\
	V &\sim N(\mu_{v}, \sigma_{v}^2),\\
	(W \given U = u, V = v) &\sim N(\alpha_{w} + \beta_{wu}u + \beta_{wv}v, \sigma_w^2),\\
	(Z \given W = w) &\sim N(\alpha_z + \beta_{zw} w, \sigma_z^2),\\
	(X \given Z = z, V = v) &\sim N(\alpha_x + \beta_{xz} z + \beta_{xv} v, \sigma_x^2),\\
	(Y \given X = x, U = u) &\sim N(\alpha_y + \beta_{yx} x + \beta_{yu} u, \sigma_y^2).
	\end{aligned}
	\end{equation}
	Here all the parameters $\mu_{\cdot}$, $\alpha_{\cdot}$, $\beta_{\cdot\cdot}$, and $\sigma_{\cdot}$ are unknown, and $U$ and $V$ are unobserved confounders. Our observational model needed for estimating $P(Y \given \doo(X = x))$ is
	\begin{equation} 
	\label{eq:cm}
	\begin{aligned}
	W &\sim N(a_{w}, s_w^2),\\
	(X \given Z = z, W = w)&\sim N(a_x + b_{xz} z + b_{xw} w, s_x^2),\\
	(Y \given X = x, Z = z, W = w) &\sim N(a_y + b_{yx} x + b_{yz} z + b_{yw} w, s_y^2),
	\end{aligned}
	\end{equation}
	where parameters $a_{\cdot}$, $b_{\cdot\cdot}$, and $s_{\cdot}$ are unknown and have to be estimated from the data.
	
	Now using equations of model~\eqref{eq:cm} to equation~\eqref{eq:dag3} yields
	\begin{equation} 
	\label{eq:eycm}
	\begin{aligned}
	E(Y \given  \doo(X = x)) &= 
	a_y + \frac{b_{yw}s_x^2}{b_{xw}^2 s_w^2+s_x^2}a_w - \frac{b_{yw}b_{xw}s_w^2}{b_{xw}^2 s_w^2+s_x^2}a_x
	+ \left(b_{yx} + \frac{b_{yw} b_{xw} s_w^2}{b_{xw}^2 s_w^2+s_x^2}\right)x\\
	&\quad+ \left(b_{yz} -\frac{b_{yw} b_{xw} s_w^2}{b_{xw}^2 s_w^2+s_x^2}b_{xz}\right)z
	\end{aligned}
	\end{equation}
	and
	\begin{equation*}
	\begin{aligned}
	Var(Y \given  \doo(X = x)) &= \frac{b_{xw}^2 s_y^2 s_w^2+s_x^2 \left(b_{yw}^2 s_w^2+s_y^2\right)}{b_{xw}^2 s_w^2+s_x^2}.
	\end{aligned}
	\end{equation*}
	While the variance $Var(Y \given  \doo(X = x))$ does not contain the \z{} $Z$, the expected value $E(Y \given  \doo(X = x))$ does, which is not surprising since $P(Y \given \doo(X = x)) = P(Y \given \doo(X = x, Z = z))$ in the graph of \autoref{fig:dagC}. Also, if the interest is only in the difference $E(Y \given \doo(X = x+1)) - E(Y \given \doo(X = x))$ then the effect of \z{} cancels out in this linear case.
	
	As our model is linear-Gaussian, we can apply path analysis \citep{wright1934} to our causal graph (see, e.g., \citet{Pearl2013} for examples) to find the marginal covariance matrix $\Sigma$ for $(Y, X, Z, W)$ (shown in Appendix). From $\Sigma$, using the properties of multivariate normal distribution, we can obtain the theoretical formulas for the parameters $(a_{\cdot}, b_{\cdot\cdot}, s_{\cdot})$ of model~\eqref{eq:cm} in terms of the true parameters $(\mu_{\cdot}, \alpha_{\cdot},\beta_{\cdot\cdot},\sigma_{\cdot})$ of the causal graph (see Appendix for details). Plugging these into equation~\eqref{eq:eycm}, we obtain
	\begin{equation*}
	\begin{aligned}
	E(Y \given  \doo(X = x)) &= \alpha_y + \beta_{yu}\mu_u + \beta_{yx} x,
	\end{aligned}
	\end{equation*}
	and
	\begin{equation*}
	\begin{aligned}
	Var(Y \given  \doo(X = x)) = \beta_{yu}^2\sigma_u^2 - 2\beta_{wu}\beta_{zw}\beta_{xz}\beta_{yx}\beta_{yu}\sigma_u^2 + \sigma_y^2.
	\end{aligned}
	\end{equation*}
	As expected, given the true causal model and its known parameters, the effect of the \z{} cancels out. Nevertheless, with a finite dataset, the estimate of expected value~\eqref{eq:eycm} depends on $z$, unless $\left(b_{yz} -\frac{b_{yw} b_{xw} s_w^2}{b_{xw}^2 s_w^2+s_x^2}b_{xz}\right)$ happens to estimate to zero.
	
\section{Simulation experiments}
\label{sec:simulations}
	
	\subsection{Binary model}
	\label{sec:bernoulli-model}
	
	We will now illustrate different choices for the \z{} with nonparametric estimation of $P(Y \given \doo(X = x))$ in a case where all variables are binary. Consider Bernoulli variables defined in accordance with \autoref{fig:dagC} as
	\begin{equation}
	\label{eq:bernoullidgm}
	\begin{aligned}
	V & \sim \text{Bernoulli}(0.5),\\
	U & \sim \text{Bernoulli}(0.5),\\
	(W \given V = v, U = u) & \sim \text{Bernoulli}(0.4 u + 0.4 v),\\
	(Z \given W = w) & \sim \text{Bernoulli}(0.4 + 0.4 w),\\
	(X \given Z = z, V = v)& \sim \text{Bernoulli}(0.4 z + 0.4 v),\\
	(Y \given X = x, U = u)& \sim \text{Bernoulli}(0.4 x + 0.4 u),
	\end{aligned}
	\end{equation}
	where $V$ and $U$ correspond to the bidirected edges between $X$ and $W$ and between $W$ and $Y$, respectively. By solving equation~\eqref{eq:dag3} analytically we obtain
	\[
	P(Y = y\given \doo(X = x)) = (0.2 + 0.4 x)^y (0.8 - 0.4 x)^{1-y}, \quad y \in \{0,1\},
	\]
	which does not depend on $z$. However, in practice when $U$ and $V$ are unobserved and we need to estimate distributions $P(W)$, $P(X \given  z, w)$, and $P(Y \given  x, z, w)$ from finite data, we will show how the estimate of $P(Y \given \doo(X = x))$ can depend on the chosen value $z$.
	
\begin{figure}[!ht]
		\centering
		\includegraphics[width=\textwidth]{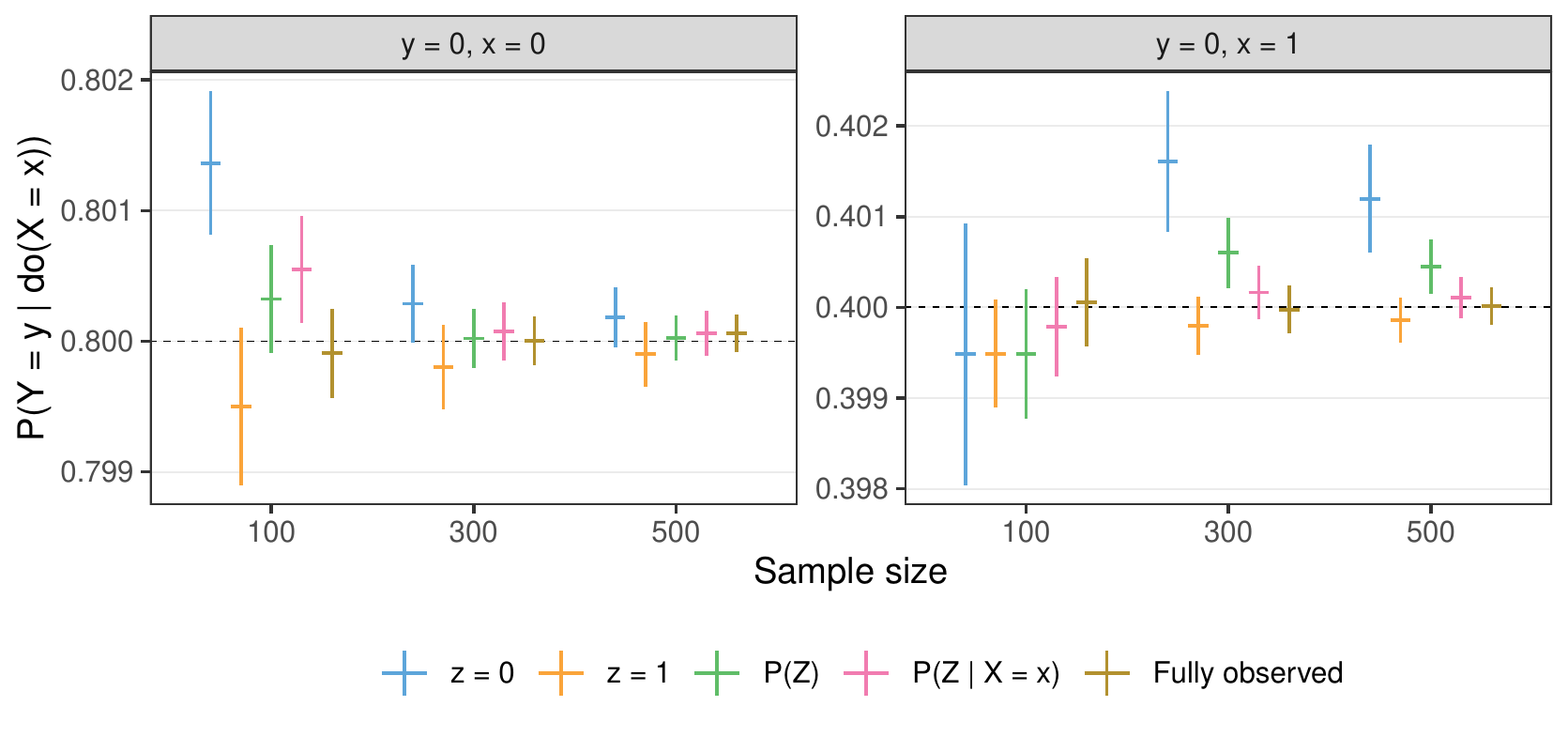} 
		\caption{Average estimates of $P(Y = 0 \given \doo(X = x))$ and $\pm 2 SE$ over 100\,000 replications for the Binary model with varying sample sizes and different strategies to account for the \z{} $Z$. The dashed lines show the true causal effects.}
		\label{fig:bernoulli}
	\end{figure}

	As an example, we simulated 100\,000 datasets according to model~\eqref{eq:bernoullidgm} with sample sizes 100, 300, and 500, and estimated $P(Y\given \doo(X = x))$ nonparametrically with various methods for dealing with the \z. Besides fixing $z$ to zero or one and using the estimator~\eqref{eq:plugin1}, we also computed a weighted average of these estimates, where the weights were based on the estimated marginal distribution $P(Z)$, or the conditional distribution $P(Z \given x)$, corresponding to the estimator~\eqref{eq:plugin2}. With sample sizes 100 and 300, there were some cases where fixing $z$ to zero or one lead to undefined probabilities (for example, there were no observations for which $x = 0$, $z = 0$ and $w = 0$). In these cases (about $10\%$ of cases with sample size 100, and less than $0.1\%$ with sample size 300), the entire replication was discarded. \autoref{fig:bernoulli} shows the results of the simulation. We see that using the fixed values with $z=0$ or $z=1$ leads to some bias. The weighted average estimators perform better, and the one based on $P(Z \given x)$ outperforms the estimator based on $P(Z)$. Overall, we are not far off from the ground truth in this simple setting.
	
	\subsection{Linear-Gaussian model}
	\label{sec:simulation_experiment}

	Now consider a model based on equation~\eqref{eq:dgm} with
	\begin{equation}
	\label{eq:lineargaussian}
	\begin{aligned}
	U &\sim N(1, 1),\\
	V &\sim N(1, 1),\\
	(W \given U = u, V = v) &\sim N(1 + u + v, 1),\\
	(Z \given W = w) &\sim N(1 + w, 1),\\
	(X \given Z = z, V = v) &\sim N(1 + z + v, 1),\\
	(Y \given X = x, U = u) &\sim N(1 + x + u, 0.01).
	\end{aligned}
	\end{equation}
	We will compare various approaches to account for the trapdoor variable $Z$. Instead of the nonparametric approach used in Section~\ref{sec:bernoulli-model}, we now switch to parametric Bayesian modelling. For comparative purposes, we use uniform priors for all unknown model parameters ($a_{\cdot}$, $b_{\cdot\cdot}$, and $s_{\cdot}$ in equation~\eqref{eq:cm}). As stated in Section~\ref{sec:estimation}, the Bayesian approach takes into account the uncertainty in $P(Y\given \doo(X = x))$ due to parameter estimation by integrating over the posterior distribution of the parameters and avoids the plug-in bias due to the nonlinearity of equation~\eqref{eq:eycm}. 
	
	For model estimation, we wrote a Stan model \citep{Stan, rstan} which simultaneously estimates all unknown model parameters and $E(Y \given \doo(X = x))$ using MCMC. We estimate the true expected causal effect $E(Y \given \doo(X = x))$ using the causal graph with observed confounders, with $x \in \{0, 3, 6, 9\}$, and compare it to the estimates obtained by six different methods:
	
	\begin{enumerate}[(a),leftmargin=*]
		\itemsep=0em
		\item Fix the trapdoor variable $Z$ to the constant $z=0$ in equation~\eqref{eq:eycm}.
		\item Marginalize over $P(Z)$ i.e., in addition of estimating model~\eqref{eq:cm} also estimate $P(Z)$, so that the estimator~\eqref{eq:eycm} uses $z = E(Z)$ at each MCMC iteration.
		\item As above, but estimate $P(Z \given  x)$ and use $z = E(Z \given x)$.
		\item Use a constraint $b_{yz} = (b_{xz}b_{yw} b_{xw} s_w^2)/(b_{xw}^2 s_w^2+s_x^2)$ so that the contribution of $z$ is fixed to zero.
		\item Fit a structural equation model (SEM), a common approach for linear-Gaussian causal modelling \citep{Kline2011}. We used the \texttt{R} package \texttt{lavaan} \citep{lavaan} for this purpose.
		\item Use the composition of weighting operators (CWO) \citep{jung2020}, which applies weighted regression to estimate functions of interventional distributions for arbitrary identifying functionals.
	\end{enumerate}
	Based on model~\eqref{eq:lineargaussian} we have $E(X) = 6$, $E(Z) = 4$, and $E(Z \given  X = 0) = 0.25$, $E(Z \given  X =3) = 2.1255$, $E(Z \given  X =6) = 4$ and $E(Z \given  X =9) = 5.875$. 
	
\begin{figure}[!ht]
		\includegraphics[width=\textwidth]{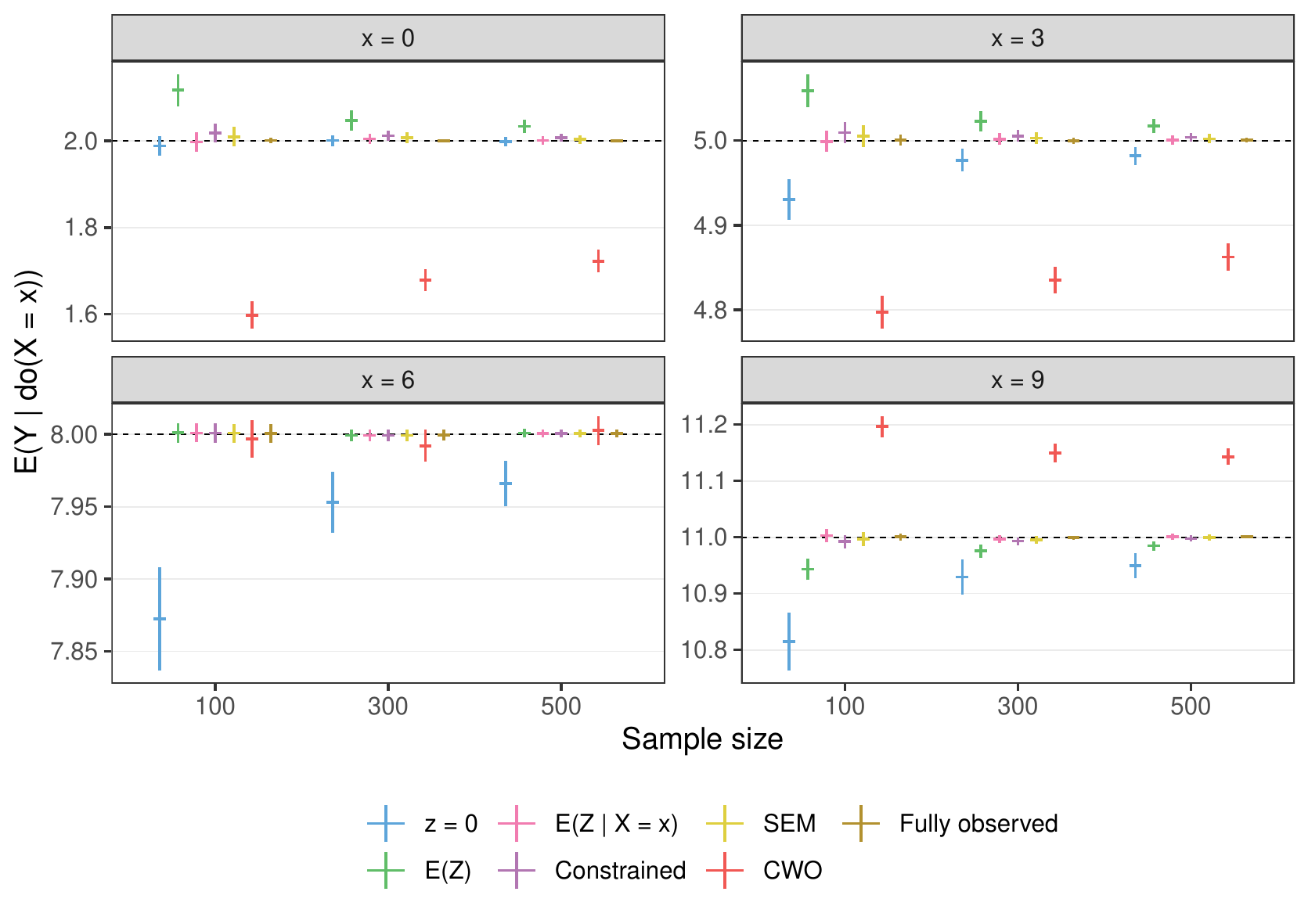}
		\caption{Average estimates of $E(Y \given \doo(X = x))$ and $\pm 2 \textrm{SE}$ over 1\,000 replications with sample sizes 100, 300 and 500 and different choices for taking the trapdoor variable into account in the simulation study of Section~\ref{sec:simulation_experiment}. The dashed lines show the true expected values.}
		\label{fig:gaussian-analytical}
	\end{figure}

	We sampled 1\,000 replications of varying sample size from model~\eqref{eq:lineargaussian} and for each replication estimated $E(Y \given \doo(X = x))$ using a single MCMC chain with 10\,000 post-warmup iterations and averaging over the iterations (thus taking account the uncertainty from parameter estimation), except for the SEM and CWO approaches which are based on the maximum likelihood estimates. \autoref{fig:gaussian-analytical} shows how, despite theoretical equivalence, results depend heavily on the chosen strategy. Perhaps the most natural choices of $z=0$ and $z=E(Z)$ are prone to bias which depends on the value $x$, but the case where $z$ is adapted based on $x$ performs well. The SEM approach, which explicitly models the covariance structures between variables, also performs well due to the structure of the graph (see \citet{sems2018} for more details), but is not applicable in more general graphs with non-Gaussian or nonlinear equations. The CWO method shows considerable bias when $x \neq E(X)$.

	\subsection{Analytical vs. Monte Carlo approach for the linear-Gaussian model}
	\label{sec:monte-carlo-example}
	
\begin{figure}[!ht]
		\includegraphics[width=\textwidth]{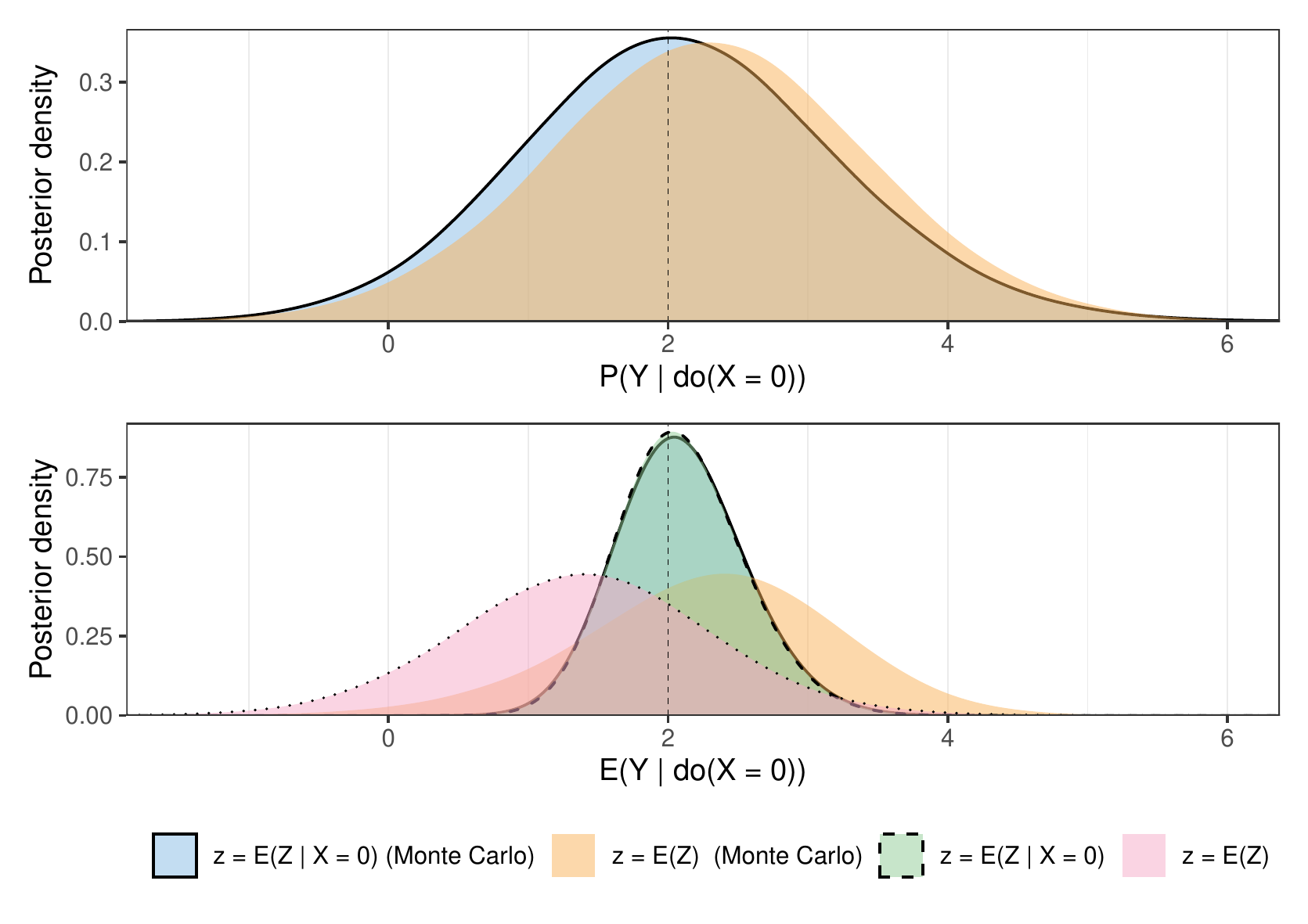} 
		\caption{Posterior predictive distribution $P(Y \given \doo(X = 0))$ and posterior distribution of $E(Y \given \doo(X = 0))$ of the linear-Gaussian model for different estimation methods in the Monte Carlo simulation of Section~\ref{sec:monte-carlo-example}.}
		\label{fig:gaussian-montecarlo}
	\end{figure}

	Consider again model~\eqref{eq:lineargaussian} but now with
	\begin{equation*}
	\begin{aligned}
	(X \given Z = z, V = v) &\sim N(1 + z + v, 0.01),
	\end{aligned}
	\end{equation*}
	i.e., smaller measurement error in $X$, leading also to a narrower density of $P(X \given z, w)$. Because of this, a poor choice for the value of the trapdoor variable can cause most of the normalized weights $\bar \gamma$ in Algorithm~\ref{algo:mc} to be close to zero, leading to inefficient Monte Carlo sampling. As an example, we simulated one replication of size $n=100$ from this model, and estimated the posterior distribution of $E(Y \given \doo(X = 0))$, both using the analytical formula and Monte Carlo with $N = 500$. For MCMC, we ran 4 chains with a total of 100\,000 post-warmup iterations. In addition, at each MCMC iteration we sampled one replication from the posterior predictive density $P(Y \given \doo(X = 0))$ from the weighted Monte Carlo sample. \autoref{fig:gaussian-montecarlo} shows the posterior distributions. We can see that with suitable $z = E(Z \given  X = 0)$ both analytical and Monte Carlo methods give approximately equal results. A poor choice of $z=E(Z)$ biases results compared to the analytical solution (which is also biased) as can be seen from the discrepancy between the posterior distributions of $E(Y \given \doo(X = 0))$ based on Monte Carlo and analytical approaches. With $N = 500$, the average MCSE (over posterior samples) was 0.08 for $z = E(Z \given  X = 0)$ and 0.22 for $z=E(Z)$. The additional variation due to the Monte Carlo sampling can inflate the posterior distribution of $E(Z \given  X = 0)$, but here with $N = 500$ the proportion of MCSE of the total posterior uncertainty is negligible: when using $z = E(Z \given X = 0)$, the posterior standard deviations of analytical and Monte Carlo methods were 0.40 and 0.41, respectively.
	
	\subsection{Non-Gaussian model}
	\label{sec:nongaussian}

	In \autoref{sec:real-data-example} we study the causal effect of education level on income using real data. Here we perform a simulation experiment where we emulate the real-data case using the assumed graph of the real-data case and define the data generating process so that it reflects the nature of the variables in \autoref{sec:real-data-example}. We use graph of \autoref{fig:fsd} where we obtain the following formula for $P(Y \given \doo(X = x))$:
	\begin{equation}
	\label{eq:fsd}
	\sum_{s,g} P(g)P(s)\frac{\sum_{w} P(Y\given x, z, w, s, g)P(x \given z,w,s,g)P(s \given w)P(w)}{\sum_{w} P(x \given z,w,s,g)P(s\given w)P(w)}.
	\end{equation}

    Let variables $U_1$, $U_2$, and $U_3$ correspond to the confounders between $W$ and $Y$, $W$ and $X$, and $W$ and $S$ respectively. For our simulation purposes we define a following data generating process:
	\begin{equation*}
	\label{eq:nongaussian}
	\begin{aligned}
	U_1 &\sim N(0, 1),\\
	U_2 &\sim N(0, 1),\\
	U_3 &\sim N(0, 1),\\
	G &\sim \text{Bernoulli}(0.5)\\
	(S \given U_3 = u_3) & \sim N(36 + 3 u_3, 25)\\
	(W \given U_1 = u_1, U_2 = u_2, U_3 = u_3) &=  \begin{cases}
	1 & \text{if $u_1 + u_2 + u_3 \leq -1.1$}\\
	2 & \text{if $-1.1 < u_1 + u_2 + u_3 \leq 1.9$}\\
	3 & \text{otherwise}
	\end{cases}\\
	(Z \given W = w, S = s, G = g) &\sim \text{Beta}(\mu_z \phi_z, (1 - \mu_z)\phi_z),\\
	(X \given Z = z, U_2 = u_2, S = s, G = g) &\sim \text{SM}(-0.5 g + 0.04 s + 13.5 z + 2 u_2, (12.5, 14)),\\
	(Y \given X = x, U_1 = u_1, S = s, G = g) &\sim \text{Gamma}(10000, 10000 / \mu_y),
	\end{aligned}
	\end{equation*}
	where 
	\begin{equation*}
	\begin{aligned}
	\mu_z &= \exp(-1.2 + 0.4 g + 0.05 s + 0.1 \textrm{I}(w = 2) + 0.3 \textrm{I}(w = 3)), \\
	\phi_z &= \exp(2.2 + 0.2 g), \\
	\mu_y &= \exp(9.3 + 0.02 s - 0.5 g + 0.2 \textrm{I}(x = 1) + 0.5 \textrm{I}(x = 2) + 0.4 u_1),
	\end{aligned}
	\end{equation*}
	
	 and $\text{SM}(\eta, \tau)$ is a sequential model \citep{Tutz1990, burkner2019} with a linear predictor $\eta$ and a threshold vector $\tau$.

	Compared to the linear-Gaussian experiment of \autoref{sec:simulation_experiment}, here additional difficulties arise due to the fact that in addition to the unknown parameters $\theta$, our distributional assumptions for the terms in equation \eqref{eq:fsd} are not correct. For example, while $(Y \given X = x, U_1 = u_1, S = s, G = g)$ is Gamma distributed by definition, $(Y \given X = x, S = s, G = g, W = w, Z = z)$ might not be. This can naturally bias our causal estimates, but the question remains whether different choices for the trapdoor variable affect the bias or precision of the causal estimates.
	
	For the terms in \eqref{eq:fsd}, we assume a Gamma distribution for $Y$, and model its expected value given other variables via log-link, assuming a monotonic effect \citep{burkner2018} of $X$ and $W$. We treat $X$ and $W$ as ordinal variables and model them with ordered logistic regression and a sequential model with a logit-link \citep{Tutz1990, burkner2019} respectively. We use a normal distribution for $S$, and a Bernoulli distribution for $G$. 
	
	We simulated 1\,000 replications of sample size 500 from this model, and estimated the posterior mean of $P(Y \given \doo(X = x))$ using the four different strategies for the trapdoor variable: $Z \sim P(Z)$, $Z \sim P(Z \given S = s^i, G = g^i)$, $Z \sim P(Z \given S = s^i, G = g^i, X = x)$, and $Z \sim P(Z \given X = x)$. We use a Beta distribution for $Z$ in all cases by modelling the expected value via a logit-link and precision via a log-link. Based on the data generating process, the true causal effects are 19\,690, 24\,049, and 32\,463 for $x=0,1,2$ respectively. The root mean square error (RMSE) and bias of our causal estimates compared to ground truth are shown in \autoref{table:nongaussian-results}. We see that RMSE increases with respect to the intervention variable $X$ and the differences between trapdoor strategies are relatively small except when $x=2$, where conditioning on the intervention variable results in approximately 20\% smaller RMSE than when using the marginal distribution of the trapdoor variable or when conditioning only with covariates $S$ and $G$. The differences in bias estimates are within the Monte Carlo error in the case of $X=0$. With $X=1$, the trapdoor strategies without conditioning on $X$ results in positive bias whereas conditioning on $X$ leads to a bias of similar magnitude but in a different direction. For $X=2$, the bias is substantially larger with trapdoor strategies not involving $X$. In this simulation experiment, conditioning on the covariates $\+ B$ does not improve the estimates. 

\begin{table}[h]
	\caption{RMSEs and biases of the estimates for the causal effect of $X$ on $Y$ in the non-Gaussian simulation experiment using four different strategies for sampling the trapdoor variable $Z$, and the true data generating process with observed confounders. MCSEs (computed with \texttt{simhelpers} R package \citep{simhelpers}) were between 0.4 and 2 for the RMSE estimates, and between 14 and 62 for the bias estimates, increasing with respect to $X$ in both cases.}
	\label{table:nongaussian-results}
	\centering
	\begin{tabular}{l|rrr|rrr}
		&		\multicolumn{3}{c|}{RMSE}	&		\multicolumn{3}{c}{Bias} \\
		\hline
		Trapdoor strategy & X = 0 & X = 1 & X = 2 & X = 0 & X = 1 & X = 2 \\
		\hline
		$P(Z \given x, s, g)$ & 528 & 875 & 1\,770 & 121 & -149 & 209\\
		$P(Z \given x)$       & 528 & 890 & 1\,729 & 123 & -233 & 4 \\
		$P(Z \given s, g)$    & 540 & 914 & 2\,238 & 106 & 110  & 1028\\
		$P(Z)$                & 544 & 912 & 2\,173 & 114 & 101  & 942 \\
		Data generating process                   & 443 & 541 & 734    & -3  & -3   & -3 \\
	\end{tabular}
\end{table}

\section{Causal effect of education on income}
	\label{sec:real-data-example}
	
	As an example with real data, we analyse Life Course 1971--2002 dataset from Finnish Social Science Data Archive \citep{fsd}. This longitudinal study consists of life courses of 634 Finnish children born in 1964--1968 in Jyv\"askyl\"a, Finland. In the early 1970s, when the children were aged between three and nine, the Illinois Test of Psycholinguistic Abilities (ITPA) was used to test their verbal intelligence. Participants were then followed up, with further information on their life events gathered in 1984, 1991, and 2002. While the dataset has been used in various studies regarding the intelligence and school achievement \citep[see, e.g.,][]{Kuusinen1988}, we use this data to study the causal effect of education level (secondary or less, lowest/lower tertiary, or higher tertiary) on yearly income (euros, in 2000). Note that due to the limited geographical scope of the data, the participants do not necessarily form a fully representative sample of the corresponding birth cohorts in Finland.
	
	In addition to earned income $Y$ and education level $X$, we have information on the GPA from primary school $Z$, socioeconomic status (SES) of the parents $W$ (low, middle, high), gender $G$, and the ITPA score $S$ (the General Language composite variable, a combination of the subtests). Out of the 634 participants in the full dataset, we use 509 participants with fully observed aforementioned variables and avoid considerations related to missing data. The assumed causal graph is shown in \autoref{fig:fsd}, and as before, all analysis are based on the assumption that this graph is correct. We use same distributional assumptions as in our simulation experiment in \autoref{sec:nongaussian}, after transforming the original scale of GPA from 4--10 to 0--1).
		
	It could be argued that there should be direct arrows from parental SES to child's education level and income. However, in Finland it is likely that the effect of SES to education is strongly mediated by the child's school achievement (variable $Z$ in our graph) \citep{Acacio-Claro}. Also, the socioeconomic status in our data was coded based on the occupations of the parents, which in turn depends on their education. We thus assume that the unobserved confounder between $W$ and $X$ contains, among other things, the education levels of the parents. Similarly, there are studies suggesting that in Finland, after accounting for a child's education, the effect of family income on children's income is low \citep{Osterbacka}. Finally, the intergenerational mobility of education and income are among the highest in Finland (and Nordic countries in general) compared to other countries \citep{Pfeffer, bjorklund}. This suggests that even if there is a direct arrow from $W$ to $X$ or $Y$, these direct effects are likely negligible. Nevertheless, it is, of course, possible (and perhaps likely) that our causal graph is an oversimplification of the complex causal mechanisms related to education and income.
	
	\begin{figure}[!ht]
		\includegraphics[width=\textwidth]{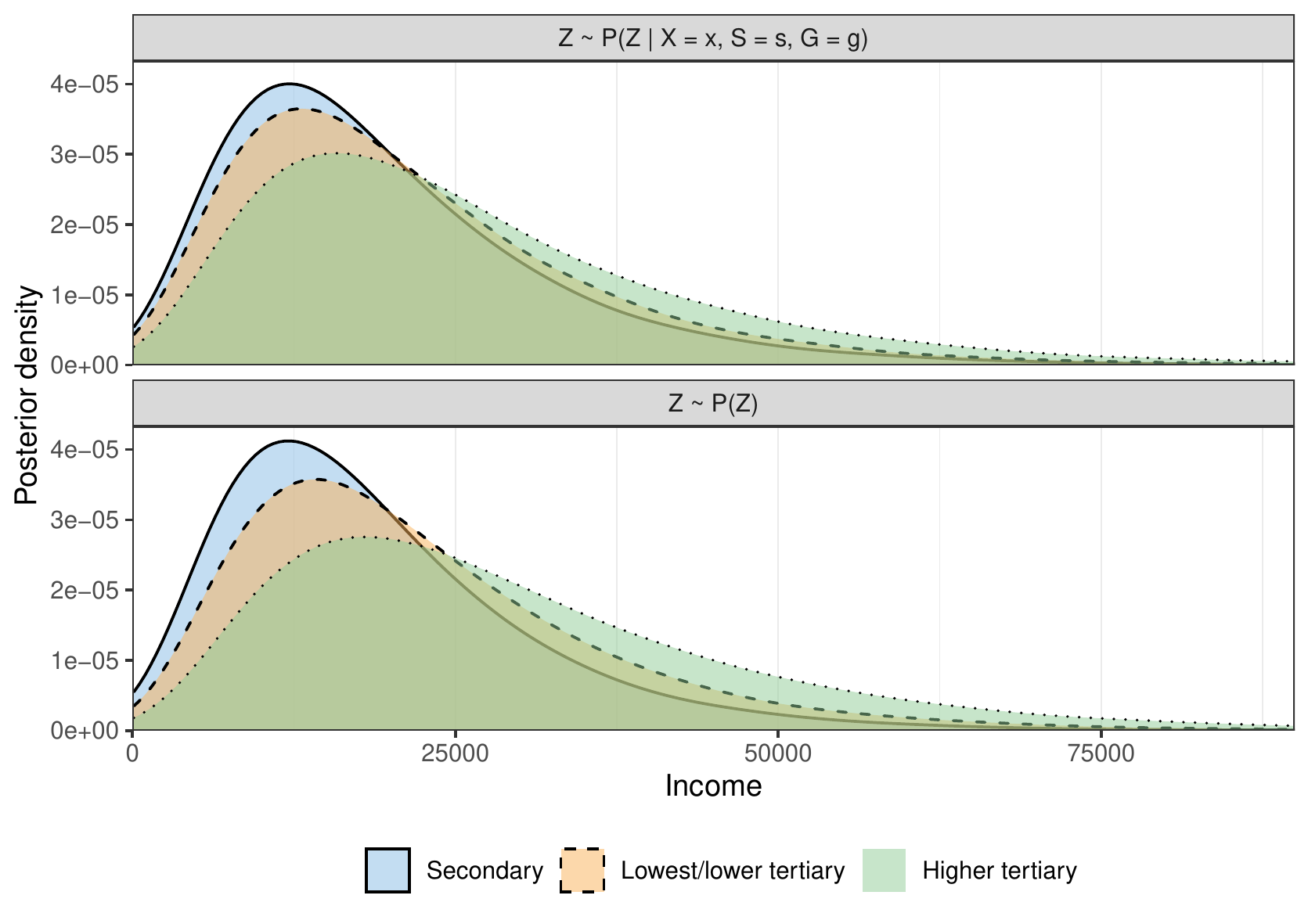} 
		\caption{Posterior predictive distribution $P(Y \given \doo(X = x))$ for the income model using the estimators with the conditional \z{} (upper panel) and the marginal \z{} (lower panel).}
		\label{fig:fsd-posterior}
	\end{figure}
	We used four chains with total of 100\,000 post-warmup iterations and $N=M=250$, leading to MCSE around 50--80, depending on $x$ with both trapdoor strategies.
	
	\autoref{table:fsd-posterior-summary} shows the posterior summary statistics for $median(Y \given \doo(X = x))$ and $E(Y \given \doo(X = x))$ for all education groups. We see a clear discrepancy between the estimates based on the conditional \z{} $Z \sim P(Z \given  x, s, g)$ and the marginal \z{} $Z \sim P(Z)$. Finally, \autoref{fig:fsd-posterior} shows the full posterior distributions $P(Y \given \doo(X = x))$. The two strategies for the trapdoor variable give somewhat different results: When conditioning the trapdoor variable grade on the education level the interventional distributions between educational levels differ more than with sampling the trapdoor variable from $P(Z)$. For example, we estimate the mean annual income for the highest education level as 29\,500 euros using the conditional \z{} compared to 26\,600 euros when using the marginal \z{}. On the basis of the simulation results of Section~\ref{sec:nongaussian}, we prefer the estimates with $Z \sim P(Z \given x, s, g)$. Unsurprisingly, obtaining a higher education level has a positive causal effect on income. Our causal estimates also differ somewhat from the observed incomes in different education groups, with the observed median incomes being (from lowest to highest) 17\,700, 19\,300, and 26\,800 euros, and the mean incomes 18\,700, 21\,800, and 28\,700 euros.
		
\begin{table}[h]
	\caption{Posterior mean, median, and their posterior standard deviations for the causal effect of education on income, in hundred euros.}
	\label{table:fsd-posterior-summary}
	\centering
	\begin{tabular}{llrrrr}
		Education level & Trapdoor strategy & Mean & Median & SD(Mean) & SD(Median) \\
		\hline
		secondary or less      & $P(Z)$                 & 195 & 166 & 8  & 6  \\
		secondary or less      &  $P(Z \given x, s, g)$ & 187 & 161 & 7  & 6  \\
		lowest/lower tertiary  & $P(Z)$                 & 216 & 183 & 12 & 11 \\
		lowest/lower tertiary  &  $P(Z \given x, s, g)$ & 223 & 192 & 13 & 11 \\
		higher tertiary        & $P(Z)$                 & 266 & 226 & 21 & 18 \\
		higher tertiary        &  $P(Z \given x, s, g)$ & 295 & 253 & 22 & 19 \\
	\end{tabular}
\end{table}

	\section{Conclusion}
	\label{sec:discussion}
	
	We have shown how it is possible to estimate causal effects when the back-door and front-door adjustments are not applicable and highlighted the potential issues related to the application of theoretical identifying formulas to finite datasets. Our real-data example on the effect of education on income illustrates how Bayesian causal inference can be applied to complex causal graphs with multiple types of variables, and how trapdoor variables can have a substantial effect on the estimated causal effects. 
	
	The basic structure that leads to implicit functional constraints and trapdoor variables was presented in the graph of \autoref{fig:dagC}. This graph has been considered in the literature earlier but according to our knowledge, it has not been fully analysed from the viewpoint of estimation. This basic structure can be extended in several directions so that the causal effect is still identifiable while retaining the implicit functional constraint. The graph for the Life Course 1971--2002 study (\autoref{fig:fsd}) is an example of such an extension. There are other interesting graphs such as those shown in \autoref{fig:extras} with similar or even more complex identifying functionals which exhibit the problems related to trapdoor variables.
	
	Determining the optimal method to account for the presence of a trapdoor variable remains a challenging problem. The trapdoor bias exhibited for small sample sizes depends on the (possibly parametric) assumptions about the causal model as well as the properties of the estimators used for the conditional distributions appearing in the identifying functional. Even in the simple case of the linear model with a single trapdoor variable, the optimal method is not apparent. It may also be the case that minimizing the trapdoor bias might result in a large variance of the causal effect estimator necessitating further considerations about the estimation problem at hand.
	
	Our simulation experiments in  Bernoulli and linear-Gaussian cases illustrated how choosing the value of the \z{} can have substantial effect in estimating the interventional distribution $P(Y \given \doo(X = x))$. Our results suggest that a good default for the \z{} should be based on its conditional distribution given the interventional variable $X$. On the other hand, for nonlinear and non-Gaussian models the effect of the \z{} can have nonlinear effects on $P(Y \given \doo(X = x))$ which can further bias the results. Therefore as a general sensitivity check, we recommend computing the causal effect using various strategies to account for the \z{}s and reporting how sensitive the results are with respect to these strategies.
	
	\section{Acknowledgements}
	
	This work belongs to the thematic research area ``Decision analytics utilizing causal models and multiobjective optimization'' (DEMO) supported by Academy of Finland (grant number 311877). We thank Yonghan Jung for providing example codes for the CWO method, and Satu Helske for providing insights on the intergenerational mobility.
		
	\appendix
	
	\section{Theoretical parameter estimates for the linear-Gaussian model of Section~\ref{sec:linear_gaussian_analytical}}
	\label{sec:appendix}
	
	Marginal covariance matrix $\Sigma$ of $Y, X, Z$ and $W$ can be found by applying path analysis in the graph of \autoref{fig:dagC}, which leads to
	\[
	\resizebox{1.0\hsize}{!}{$
		\Sigma = \begin{pmatrix}
		\sigma_{yy} & \cdot & \cdot & \cdot \\
		\sigma_{xx}\beta_{yx} + \sigma_{uu}\beta_{xz}\beta_{zw}\beta_{wu}\beta_{yu} & \sigma_{xx} & \cdot & \cdot \\
		\sigma_{zz}\beta_{xz}\beta_{yx} + \sigma_{uu}\beta_{zw}\beta_{wu}\beta_{yu} + \sigma_{vv}\beta_{zw}\beta_{wv}\beta_{xv}\beta_{yx} &
		\sigma_{zz}\beta_{xz} + \sigma_{vv}\beta_{zw}\beta_{wv}\beta_{xv} &
		\sigma_{zz} & \cdot \\
		\sigma_{ww}\beta_{zw}\beta_{xz}\beta_{yx} + \sigma_{uu}\beta_{wu}\beta_{yu} + \sigma_{vv}\beta_{wv}\beta_{xv}\beta_{yx} &
		\sigma_{ww}\beta_{zw}\beta_{xz} + \sigma_{vv}\beta_{wv}\beta_{xv} &
		\sigma_{ww}\beta_{zw} & \sigma_{ww}
		\end{pmatrix}.$}
	\]
	From $\Sigma$ we can obtain the following
	\begin{flalign*}
	b_y &= \begin{pmatrix}
	b_{yx} \\
	b_{yz} \\
	b_{yw}
	\end{pmatrix} = \Sigma_{1, 2:4} \Sigma_{2:4, 2:4}^{-1} & \\
	&= \begin{pmatrix}
	\beta_{yx} - \dfrac{\beta_{wu} \beta_{wv} \beta_{xv} \beta_{yu} \sigma_{u}^2 \sigma_{v}^2}{\beta_{wu}^2 \beta_{xv}^2 \sigma_{u}^2 \sigma_{v}^2 + \sigma_{w}^2 (\beta_{xv}^2 \sigma_{v}^2 + \sigma_{x}^2) + \beta_{wu}^2 \sigma_{u}^2 \sigma_{x}^2 + \beta_{wv}^2 \sigma_{v}^2 \sigma_{x}^2} \\[9pt]
	\beta_{xz}\dfrac{\beta_{wu} \beta_{wv} \beta_{xv} \beta_{yu}  \sigma_{u}^2 \sigma_{v}^2}{ \beta_{wu}^2 \beta_{xv}^2 \sigma_{u}^2 \sigma_{v}^2 + \sigma_{w}^2 (\beta_{xv}^2 \sigma_{v}^2 + \sigma_{x}^2) + \beta_{wu}^2 \sigma_{u}^2 \sigma_{x}^2 + \beta_{wv}^2 \sigma_{v}^2 \sigma_{x}^2} \\[9pt]
	\dfrac{\beta_{wu} \beta_{yu} \sigma_{u}^2 (\beta_{xv}^2 \sigma_{v}^2 + \sigma_{x}^2)}{\beta_{wu}^2 \beta_{xv}^2 \sigma_{u}^2 \sigma_{v}^2 +  \sigma_{w}^2 (\beta_{xv}^2 \sigma_{v}^2 + \sigma_{x}^2) + \beta_{wu}^2 \sigma_{u}^2 \sigma_{x}^2 + \beta_{wv}^2 \sigma_{v}^2 \sigma_{x}^2}
	\end{pmatrix}, & \\[5pt]
	b_x &= \begin{pmatrix}
	b_{xz} \\
	b_{xw} 
	\end{pmatrix} = \Sigma_{2, 3:4} \Sigma_{3:4, 3:4}^{-1} = 
	\begin{pmatrix}
	\beta_{xz} \\
	\dfrac{\beta_{wv}\beta_{xv}\sigma_v^2}{\beta_{wu}^2\sigma_u^2 + \beta_{wv}^2\sigma_v^2 + \sigma_w^2}
	\end{pmatrix}, & \\
	a_w &= \alpha_w + \beta_{wu}\mu_u + \beta_{wv}\mu_v,\\
	a_x &= \alpha_{x} + \dfrac{\beta_{xv} \mu_{v} (\beta_{wu}^2 \sigma_{u}^2 + \sigma_{w}^2) - \beta_{wv} \beta_{xv} \sigma_{v}^2 (\alpha_{w} + \beta_{wu} \mu_{u})}{\beta_{wu}^2 \sigma_{u}^2 + \beta_{wv}^2 \sigma_{v}^2 + \sigma_{w}^2}
	, & \\
	a_y &= \alpha_y - \{\beta_{wu} \beta_{yu} (\alpha_{w} \sigma_{u}^2 \sigma_{x}^2 + \alpha_{w} \beta_{x}v^2 \sigma_{u}^2 \sigma_{v}^2 + \beta_{wv} \mu_{v} \sigma_{u}^2 \sigma_{x}^2 - \alpha_{x} \beta_{wv} \beta_{x}v \sigma_{u}^2 \sigma_{v}^2) & \\
	&\quad - \beta_{yu} \mu_{u} (\beta_{wv}^2 \sigma_{v}^2 \sigma_{x}^2 + \beta_{x}v^2 \sigma_{v}^2 \sigma_{w}^2 + \sigma_{w}^2 \sigma_{x}^2)\}\,/\\
	&\qquad (\beta_{wu}^2 \beta_{x}v^2 \sigma_{u}^2 \sigma_{v}^2 + \beta_{wu}^2 \sigma_{u}^2 \sigma_{x}^2 + \beta_{wv}^2 \sigma_{v}^2 \sigma_{x}^2 + \beta_{x}v^2 \sigma_{v}^2 \sigma_{w}^2 + \sigma_{w}^2 \sigma_{x}^2), & \\
	s_x^2 &= \Sigma_{2,2} - \Sigma_{2,3:4} \Sigma_{3:4,3:4}^{-1} \Sigma_{2, 3:4}'
	=\sigma_{x}^2 + \beta_{xv}^2 \sigma_{v}^2 \dfrac{\beta_{wu}^2 \sigma_{u}^2 + \sigma_{w}^2}{\beta_{wu}^2 \sigma_{u}^2 + \beta_{wv}^2 \sigma_{v}^2 + \sigma_{w}^2}, & \\
	s_w^2 &= \Sigma_{4,4}  = \beta_{wu}^2\sigma_u^2 + \beta_{wv}^2\sigma_v^2 + \sigma_w^2.
	\end{flalign*}
	
	\bibliographystyle{rss}
	\bibliography{references}
\end{document}